\documentclass[journal,twoside,web]{ieeecolor}
\usepackage{tmi}
\usepackage{cite}
\usepackage{amsmath,amssymb,amsfonts}
\usepackage{algorithmic}
\usepackage{graphicx}
\usepackage{textcomp}
\def\BibTeX{{\rm B\kern-.05em{\sc i\kern-.025em b}\kern-.08em
    T\kern-.1667em\lower.7ex\hbox{E}\kern-.125emX}}
\usepackage{xcolor}
\usepackage{subcaption}
\usepackage{booktabs}
\usepackage{tabularx}
\usepackage{url}
\usepackage[dvipsnames]{xcolor}
\usepackage[linesnumbered,ruled,vlined]{algorithm2e}
\usepackage{array} %

\newcolumntype{C}[1]{>{\centering\arraybackslash}p{#1}}

\newcommand{\exvivo}{\emph{ex vivo }} 
\newcommand{\invivo}{\emph{in vivo }} 

\begin{document}
\title{Reference-Free 3D Reconstruction of Brain Dissection Slabs via Learned Atlas Coordinates}
\author{Lin Tian, Jonathan Williams-Ramirez, Dina Zemlyanker, Lucas J. Deden-Binder, Rogeny Herisse, Theresa R. Connors, Mark Montine, Istvan N Huszar, Lilla Z\"ollei, Sean I. Young, Christine Mac Donald, C. Dirk Keene, Derek H. Oakley, Bradley T. Hyman, Oula Puonti, Matthew S. Rosen, and Juan Eugenio Iglesias
\thanks{This work was supported by the NIH (grants RF1-MH123195, R01-AG070988, R01-EB031114, UM1-MH130981, RF1-AG080371, R21-NS138995, P30-AG062421, and K99-AG081493) and by the Lundbeck Foundation (R360–2021–39). M.S. Rosen acknowledges the support of the Kiyomi and Ed Baird MGH Research Scholar award. Istvan N Huszar and Lilla Z\"ollei are supported by NIH R01HD102616.}
\thanks{
Lin Tian (e-mail:ltian3@mgh.harvard.edu), Jonathan Williams-Ramirez, Dina Zemlyanker, Lucas J. Deden Binder, Rogeny Herisse, Istvan N Huszar, and Lilla Z\"ollei are affiliated with the Martinos Center for Biomedical Imaging (Martinos Center) at Massachusetts General Hospital (MGH) \& Harvard Medical School (HMS).
}
\thanks{
Sean I. Young is affiliated with the Martinos Center at MGH \& HMS, and the Computer Science and Artificial Intelligence Laboratory (CSAIL) at the Massachusetts Institute of Technology (MIT).
}
\thanks{
Oula Puonti is affiliated with MGH \& HMS. He is also affiliated with the Danish Research Centre for Magnetic Resonance, Centre for Functional and Diagnostic Imaging and Research, Copenhagen University Hospital-Amager and Hvidovre, Copenhagen, Denmark.
}
\thanks{
Theresa R. Connors is affiliated with Massachusetts Alzheimer's Disease Research Center at MGH \& HMS.
}
\thanks{
Mark Montine, Christine Mac Donald, and C. Dirk Keene are affiliated with the University of Washington.
}
\thanks{
Derek H. Oakley is with the Pathology Department at MGH \& HMS.
}
\thanks{
Bradley T. Hyman is with the Neurology Department at MGH \& HMS.
}
\thanks{
Matthew S. Rosen is with the Martinos Center at MGH \& HMS and with the Department of Physics at Harvard University.
}
\thanks{
Juan Eugenio Iglesias is with the Martinos Center at MGH \& HMS and the CSAIL at MIT.
}}

\maketitle

\begin{abstract}
Correlation of neuropathology with MRI has the potential to transfer microscopic signatures of pathology to \invivo scans. 
There is increasing interest in building these correlations from 3D reconstructed stacks of slab photographs, which are routinely taken during dissection at brain banks. These photographs bypass the need for \exvivo MRI, which is not widely accessible. However, existing methods either require a corresponding 3D reference  (e.g., an \exvivo MRI scans, or a brain surface acquired with a structured light scanner) or a full stack of brain slabs, which severely limits applicability. Here we propose ``RefFree'', a 3D reconstruction method for dissection photographs that does not require an external reference. RefFree coherently reconstructs a 3D volume for an arbitrary set of slabs (including a single slab) using predicted 3D coordinates in the standard atlas space (MNI) as guidance. To support RefFree's pipeline, we train an atlas coordinate prediction network that estimates the coordinate map from a 2D photograph, using synthetic photographs generated from digitally sliced 3D MRI data with randomized appearance for enhanced generalization. As a by-product, RefFree can propagate information (e.g., anatomical labels) from atlas space to one single photograph even without reconstruction. Experiments on simulated and real data show that, when all slabs are available, RefFree achieves performance comparable to existing classical methods but at substantially higher speed. Moreover, RefFree yields accurate reconstruction and registration for partial stacks or even a single slab. Our code is available at \url{https://github.com/lintian-a/reffree}.
\end{abstract}

\begin{IEEEkeywords}
Neuroimaging-pathology correlation, Joint image registration,  atlas registration
\end{IEEEkeywords}

\footnotetext{This work has been submitted to the IEEE for possible publication. Copyright may be transferred without notice, after which this version may no longer be accessible.}

\section{Introduction}
\begin{figure*}[h]
    \centering
    \includegraphics[width=0.94\linewidth]{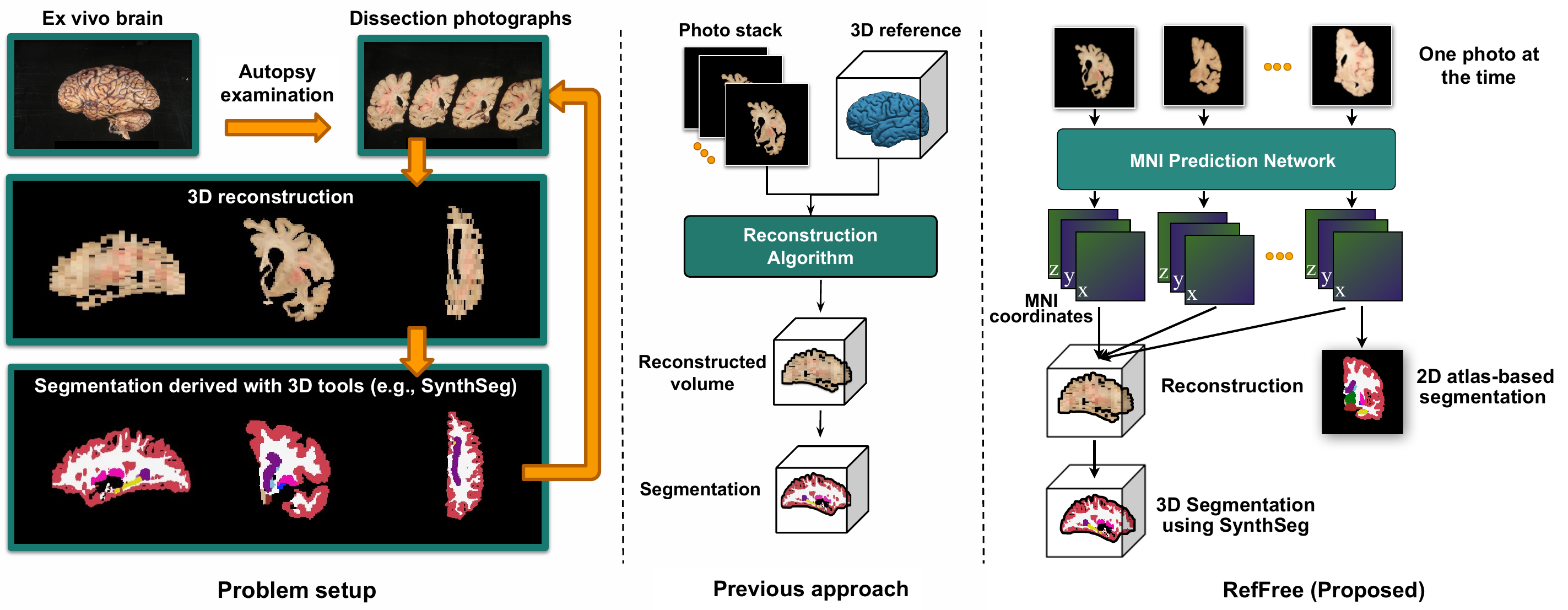}
    \caption{\textbf{Problem setup and proposed approach RefFree}. In our problem (left), we are given a stack of dissection photographs, which we aim to reconstruct into a 3D brain volume with its structures labeled. Previous methods (middle) require a reference image (e.g., a surface or \emph{premortem} MRI scan)  to guide this 3D reconstruction.
    The proposed method (right) solves this problem using a neural network that receives only the photographs as input and outputs predictions of voxel coordinates in the standard MNI space, which are fit to a transform for 3D reconstruction.}
    \label{fig:teaser}
\end{figure*}

Neuropathology is the gold standard for diagnosing neurodegenerative diseases such as Alzheimer’s disease. A typical autopsy workflow begins with a macroscopic inspection of the brain, performed on thick coronal slices (5-10~mm, Fig.~\ref{fig:teaser}), and proceeds to microscopic analysis of sections cut from selected tissue blocks. In recent years, increasing attention has been paid to correlating histological findings with \invivo or \exvivo MRI, with the goal of identifying radiographic biomarkers that reflect underlying pathology~\cite{dawe2011neuropathologic,kotrotsou2015neuropathologic,huszar2023tensor}.

Histology-MRI alignment methods commonly reconstruct a 3D histology volume from serial slices, and then perform deformable registration to MRI~\cite{pichat2018survey}. While such pipelines may achieve voxel-level correspondence in small animal brains or focal human resections, it becomes increasingly challenging for larger organs such as the human brain, where histological sections sample only centimeter-scale regions. Without a global context, localizing a single histological section within a much larger full-brain MRI volume is very difficult and unreliable. This is due to the combination of the nonlinear deformation in the histology, the missing tissue caused during acquisition, the inherent ambiguity in 2D-3D registration, and differences in image contrast and resolution.

To bridge this gap, several recent studies~\cite{huszar2023tensor,tregidgo20203d,gazula2024machine} have proposed using the coronal slab photographs as a practical intermediary. 
These photographs preserve the full anatomical context compared to a histological section and are already available in many brain banks and clinical archives. Thus, they may serve as a spatial conduit between the histological section and the MRI of a full brain, paving the way for reliable neuroimaging-pathology correlation.

A recent approach~\cite{tregidgo20203d,gazula2024machine}, publicly available in FreeSurfer~\cite{fischl2012freesurfer}, reconstructs a full stack of slab photographs by simultaneously registering the slabs to each other under the guidance of a 3D reference (e.g., an MRI, surface scan, or an atlas) using iterative optimization. This enables coherent 3D reconstruction and downstream 3D analysis. However, it assumes access to a complete stack of slab photographs, which limits its applicability when some slabs may be missing, damaged, or never imaged. In contrast, TIRL~\cite{huszar2023tensor} aligns a single slab to an \exvivo MRI given an approximate slice location, but this method does not guarantee a coherent reconstruction provided a set of slabs. Consequently, current approaches lack a unified framework that can accommodate varying levels of available input, from a full stack to a partial stack to a single slab. In addition, they rely on multi-resolution iterative optimization, which can be computationally intensive and time-consuming. Although runtime is not the primary concern in neuroimaging–pathology correlation applications, a more efficient solution is still desirable.

Here we introduce RefFree, a learning-based framework that reconstructs 3D brain volumes from coronal slab photographs without any specimen-specific reference. %
RefFree uses a 2D convolutional neural network to predict the corresponding MNI atlas~\cite{fonov2011unbiased} coordinate for each pixel in the single photograph input. Predicted coordinates from one or multiple slabs are then jointly fitted with least-squares to reconstruct a coherent 3D volume. Because the network operates independently on each image, RefFree works with a complete stack, a partial stack, or even a single photograph. As a byproduct, segmentations of brain regions can be obtained via forward projection from MNI space in near real-time. To train RefFree, we propose a synthetic data engine that generates anatomically plausible slab-like images from 3D MRI segmentations. This engine simulates variability in anatomy, cutting angle, lighting, tissue contrast, and deformation. This domain-randomized engine eliminates the need for any manually labeled photographs and enables robust transfer to real autopsy data. 

In summary, our key contributions are:
\begin{itemize}
    \item A framework for reconstructing 3D brain volumes from dissection slab photographs via coordinate prediction.
    \item A synthetic training pipeline that incorporates anatomical and acquisition variability, supporting the training of the coordinate-prediction network on slab photographs and facilitating robust generalization to real-world autopsy data.
    \item Support for reference-free reconstruction from partial or single slabs, enabling wide, retrospective applicability.
    \item Comprehensive quantitative and qualitative evaluation of RefFree on \textit{(i)}~a synthetic benchmark derived from 11 public MRI cohorts; and \textit{(ii)}~two post-mortem datasets comprising 149 single-hemisphere stacks and 26 bi-hemispheric stacks. Experiments assess coordinate-prediction accuracy, ablate synthetic-engine components, measure full-stack reconstruction fidelity, test robustness to missing slabs, and demonstrate real-time anatomical overlay on individual photographs.
\end{itemize}

\section{Related Work}
\subsection{3D Reconstruction from slice images}
Volumetric reconstruction from sparsely sampled 2D images is a long-standing challenge in medical imaging, underpinning applications in histology~\cite{pichat2018survey, casamitjana2025probabilistic}, block-face electron microscopy~\cite{wanner2015challenges}, or 3D ultrasound~\cite{huang2017review}. For histology reconstruction, two principal methodological families dominate the literature~\cite{pichat2018survey}. The first reconstructs a volume using the slices alone: consecutive sections are registered under the assumption that tissue shape varies smoothly along the stacking direction.
Without an external frame of reference, alignment errors can accumulate an artifact commonly known as drift or ``z-shift''. Furthermore,  the reconstructed volume may not remain geometrically faithful to the specimen due to the so-called ``banana effect'', (e.g., the straightening of curved structures)~\cite{malandain2004fusion}. The second family anchors the slices to a reference volume, typically a corresponding \exvivo MRI, CT, or an anatomical atlas, thus reducing drift and improving geometric accuracy.
However, as the number of available photographs decreases, registration becomes progressively more challenging. With a handful, or even a single slab, the task reduces to registering one or few planes to a 3-D volume, a problem highly sensitive to initial pose and prone to local optima during optimization~\cite{ferrante2017slice}. Most brain bank collections lack specimen-specific references, and many do not have complete photo stacks either, which limits the applicability of existing techniques. The framework proposed in this study effectively addresses these limitations. 

\subsection{Learning Correspondence via Neural Networks}
Learning correspondence via neural networks can be achieved through supervised or unsupervised paradigms.
Unsupervised methods typically rely on similarity measures between input images combined with regularization losses, and they have become the dominant strategy in learning-based medical image registration~\cite{de2017end,balakrishnan2019voxelmorph,hoffmann2021synthmorph,mok2020large,chen2022transmorph,tian2023gradicon,tian2024unigradicon}.
In contrast, supervised approaches exploit ground-truth correspondences or transformation parameters, either generated synthetically, as in optical flow networks~\cite{sun2018pwc,teed2020raft}, or obtained from conventional iterative registration methods, as in supervised medical registration frameworks~\cite{yang2017quicksilver,sokooti2017nonrigid,gopinath2024registration}. In this work, we build on the general idea of learning correspondence via supervision; however, rather than applying it to registration, we exploit the correspondences predicted by a trained network as geometric constraints to guide 3D reconstruction from 2D images.

\subsection{Domain Randomization for Medical Image Networks}
A further challenge lies in the absence of supervised data. Ideally, one would have slab photographs exhibiting broad appearance variability, each annotated with atlas coordinates. Such data, however, are not available in practice, making direct supervision on real photographs infeasible. Recent work in medical imaging addresses similar annotation gaps by training on synthetic volumes produced through domain randomization~\cite{gopinath2024synthetic}. In brain MRI, such synthetic‐to‐real strategies have improved segmentation (e.g., SynthSeg~\cite{billot2023synthseg,billot2023robust}), super-resolution and reconstruction (SynthSR~\cite{iglesias2023synthsr}), modality-agnostic foundation modeling (BrainID~\cite{liu2024brain}), and registration (SynthMorph~\cite{hoffmann2021synthmorph}). These studies randomize resolution (by resampling) and image contrast (by sampling Gaussian‐mixture models conditioned on tissue labels) and yield robust generalization to clinical MRI scans of any resolution and contrast. However, all of the above-mentioned methods are proposed for 3D structural MRIs, whose data engine, therefore, cannot be readily applied to \exvivo brain slab photographs.

\begin{figure*}[!h]
    \centering
    \includegraphics[width=0.9\linewidth, trim={1cm 0 6.5cm 0},clip]{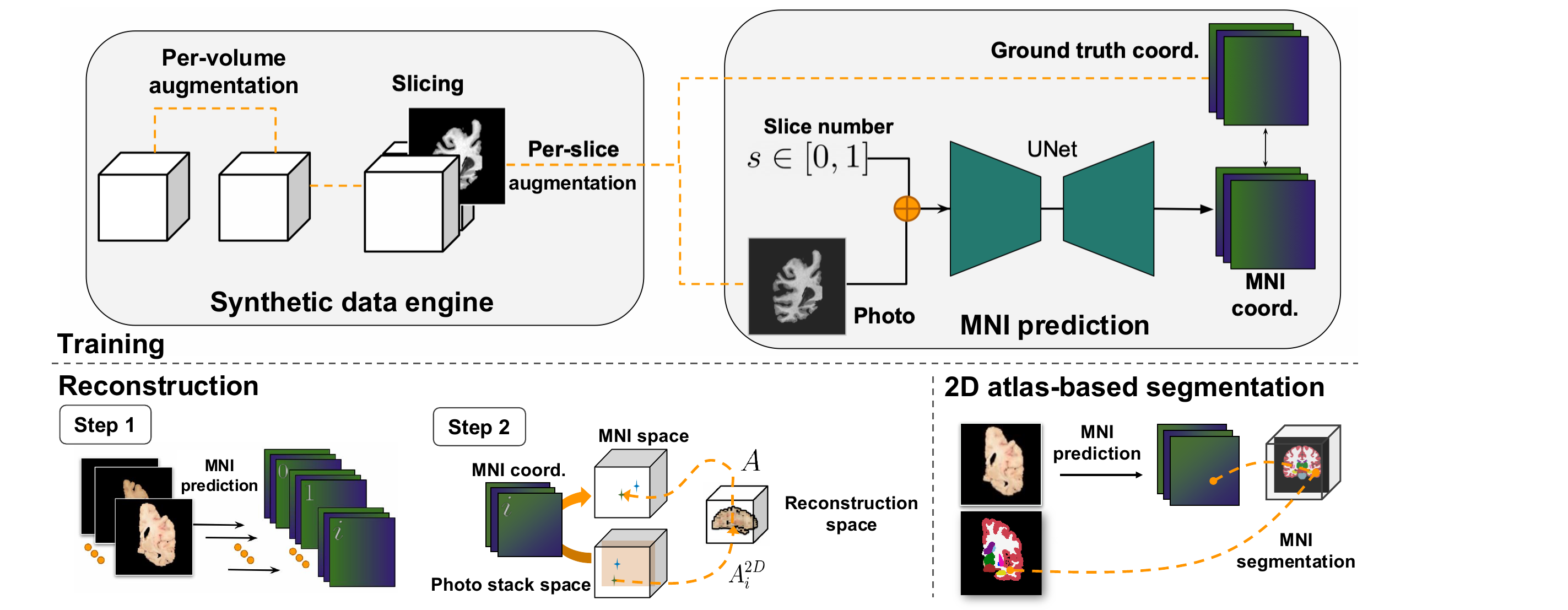}
    \caption{\textbf{Training and inference.} Top: A 2D photograph and its corresponding MNI coordinate map are generated from the 3D MRI dataset, and a U-Net is trained to predict MNI coordinates from a single photograph. Bottom left: During inference, predicted MNI coordinates for each available photograph enable analytic 3D reconstruction. Bottom right: The same predicted coordinates can be used to obtain an atlas-based segmentation map even without performing reconstruction.}
    \label{fig:pipeline}
\end{figure*}

\section{Methods}
RefFree consists of two stages (Fig.~\ref{fig:pipeline}). First, a CNN predicts the corresponding 3D atlas (MNI) coordinates for each pixel in each slab photograph (Sec.\ref{sec:network}). Because photographs are processed independently, the method naturally handles full stacks, partial stacks, or even single images. Second, given these coordinate predictions, we analytically estimate a set of transformations that jointly map the photographs into MNI space (Sec.\ref{sec:reconstruction}). This step leverages the geometric consistency across the 2D photographs to ensure a coherent, geometrically plausible 3D arrangement.

\subsection{MNI Coordinate Prediction from Dissection Photographs}\label{sec:network}

\paragraph{\textbf{Task formulation}} Let $I: \Omega_s \rightarrow \mathbb{R}$ be a grayscale 2D image defined on the normalized domain $\Omega_s: [-1,1]^2$ where $s \in [0,1]$ represents its normalized position along the posterior–anterior axis. We train a CNN $f_\theta$ parameterized by $\theta$ to predict, for each pixel, its corresponding 3D coordinate in the normalized MNI atlas space $\Omega_t = [-1,1]^3$:
\begin{equation}
f_\theta(I, s) = \hat{Y}, \quad \hat{Y}: \Omega_s \rightarrow \Omega_t.
\end{equation}
The normalized position $s$ (as defined in Eq.~\ref{equ:s_definition}) provides global context, facilitating the model's generalization across brains of varying sizes and slab thicknesses.

\paragraph{\textbf{Synthetic Data Engine for Training}} 
Preparing a real dataset with sufficiently diverse slab photographs (e.g., different cameras, illumination conditions, and cutting angles) would be extremely impractical, and obtaining gold-standard MNI coordinates would also be challenging. We instead create synthetic slab photographs from publicly available high-resolution T1-weighted MRIs. From each MRI, we derive a 3D segmentation label map $L: \Omega_v \rightarrow \mathbb{N}$ (where $\Omega_v$ is the 3D MRI volume domain) using FreeSurfer~\cite{fischl2012freesurfer}. This segmentation is augmented with additional labels for limbic structures and the claustrum, obtained through multi-atlas~\cite{iglesias2015multi} and deep learning methods~\cite{greve2021deep}. A corresponding MNI coordinate field $Y: \Omega_v \rightarrow \Omega_t$ is computed via classical nonlinear registration to the MNI atlas using NiftyReg~\cite{modat2014global}. Synthetic training samples are generated in multiple steps:

\vspace{2mm}
\noindent\textbf{Step 1: Simulation of Anatomy Pose and Cutting Pose.} The \exvivo brain could undergo various deformations and a slightly tilted slice pose during the slab cutting. To mimic such variance, we apply a random 3D affine transformation to the label map $L$ and MNI coordinate field $Y$:
\begin{equation}
    L^{\prime}=L\circ\mathcal{A}, \;\; Y^{\prime}=Y\circ\mathcal{A},
\end{equation}
where $\mathcal{A}\in Aff(3)$ includes rotation, scaling, and shear. Parameters for each transformation are independently sampled from uniform distributions with predefined parameter ranges:
\begin{equation}
    \begin{aligned}
        \theta_{\text{rot}} &\sim \mathcal{U}(-\alpha_{r}, \alpha_{r}),\\
        \theta_{\text{scale}} &\sim \mathcal{U}(1-\beta_{s}, 1+\beta_{s}),\\
        \theta_{\text{shear}} &\sim \mathcal{U}(-\gamma_{h}, \gamma_{h}).
    \end{aligned}
\end{equation}

\vspace{2mm}
\noindent\textbf{Step 2: Simulation of slabbing process.} We digitally slice the transformed volumes $L^\prime$ and $Y^\prime$ along the coronal axis. Given a sequence of $K$ slices from posterior to anterior, the normalized slice number $s_k\in[0,1]$ for the $k$-th slab is defined as
\begin{equation}\label{equ:s_definition}
    s_k=\frac{k-1}{K-1},\quad k=1,\dots,K.
\end{equation}
This coronal slicing approach is sufficient because cutting angle and tissue deformation have already been simulated in Step~1.

\vspace{2mm}
\noindent\textbf{Step 3: Slab Deformation and Framing.} In the neuropathology gross exam, the cut slabs are laid out in a tray with a camera placed on top for the dissection slab photographs. During transportation from the cutting station to the tray, deformations occur in the slab. To simulate realistic deformations occurring between slab cutting and photographing, we simulate non-rigid slab deformation by sampling a low-resolution field of 2D displacement vectors from a Gaussian distribution. Specifically, at each pixel of the deformation field, we sample each component of the displacement vector via
\begin{equation}
    \delta\sim\mathcal{N}(0,\sigma^2),\quad\sigma\sim\mathcal{U}(0,\sigma_{\text{max}}),
\end{equation}
where $\sigma_{\text{max}}$ controls the deformation magnitude. The low-resolution displacement field is bilinearly interpolated to full resolution and applied independently to each slice in the label map $L^\prime$ and its corresponding MNI coordinates $Y^\prime$. Additionally, we apply random cropping to each slice to simulate variability in slab position and photographic framing.

\vspace{2mm}
\noindent\textbf{Step 4: Synthetic Photograph from the Label Map.} Each slab’s anatomical label map is converted into a photographic image using a Gaussian mixture model (GMM). For each anatomical label $l$, mean and standard deviation intensities $\mu_l$ and $\sigma_l$ are sampled uniformly with predefined parameters:
\begin{equation}
    \mu_l\sim\mathcal{U}(\mu_{\text{min}},\mu_{\text{max}}),\quad\sigma_l\sim\mathcal{U}(\sigma_{\text{min}},\sigma_{\text{max}}).
\end{equation}
Pixel intensities are then drawn from the corresponding Gaussian distributions:
\begin{equation}
    I(x)\sim\mathcal{N}(\mu_{L(x)},\sigma_{L(x)}^2).
\end{equation}

The actual photographs could suffer from uneven illumination due to non-uniform lighting conditions. To model this phenomenon, we generate a smooth multiplicative field $\mathcal{E}$. At each voxel of $\mathcal{E}$, the multiplicative value is sampled from: 
\begin{equation}
    \log \mathcal{E}\sim\mathcal{N}(0,\sigma_e^2),\quad\sigma_e\sim\mathcal{N}(0,\sigma_{\text{illum}}),
\end{equation}
which is interpolated to full image resolution. The final synthetic photograph is obtained by multiplicative intensity modulation:
\begin{equation}
    I_{\text{final}}(x)=I(x)\cdot \mathcal{E}(x).
\end{equation}

\paragraph{\textbf{CNN Training}} Each training data consists of a synthetic slab photograph $I_i$, its corresponding normalized slice index $s_i$, and the ground-truth MNI coordinate map $Y_i$. At each minibatch iteration, we randomly select one MRI volume and run the synthetic data generation pipeline (Steps 1–4 described above). We then randomly sample N slab images from the generated set of synthetic slabs to form the training batch.

The CNN parameters are optimized by minimizing the mean absolute error between predicted and ground-truth coordinates, computed only over foreground pixels:

\begin{equation}
\mathcal{L}(\theta) = \sum_{i=1}^{N} \frac{1}{|M_i|_1}|M_i \odot (f\theta(I_i,s_i)-Y_i)|_1, \label{equ:loss}
\end{equation}
where $M_i$ is the binary foreground mask, and $\odot$ denotes element-wise multiplication. Training is performed for 100 epochs using the Adam optimizer with a learning rate of $10^{-4}$. 

\paragraph{\textbf{Implementation Details}} We tune the parameters of the synthetic data engine according to the performance of the prediction network on the validation set of the Composite Brain MRI dataset (Sec.~\ref{sec:dataset}). It yields the following parameters. In Step 1, we use $\alpha_r=15$, $\beta_s=0.2$, $\gamma_h=0.2$. In Step 3, we use $\sigma_{max}=4$. In Step 4, we use $\mu_{min}=0.02$, $\mu_{max}=0.04$, $\sigma_{min}=0.1$, $\sigma_{max}=0.6$, and $\sigma_{illum}=0.1$.

\paragraph{\textbf{Inference on Real Photographs}} 
Real photographs are first preprocessed by converting to grayscale, min-max normalization, correcting for pixel size, segmenting the foreground, and ordering from anterior to posterior, all achieved with FreeSurfer tools~\cite{gazula2024machine}.
The normalized, masked images and slice numbers are input into the CNN to predict the MNI maps. 

\subsection{3D Reconstruction}
\label{sec:reconstruction}

Our goal is to reconstruct a coherent 3D volume given the predicted MNI coordinates of a stack of slab photographs. A straightforward strategy would be to estimate a separate 3D affine transformation ${A^{3D}_i}$ for each slab by minimizing the least-squares error between the transformed pixel coordinates and the predicted MNI coordinates. However, fitting each slab independently may produce overlapping or intersecting slabs in 3D if the predicted coordinates contain inconsistencies, thereby breaking the global geometric consistency of the reconstructed volume.

We instead factor per-slab 3D transformation into a shared global 3D affine transformation $A$ and a per-slab in-plane 2D affine transformation $A^{2D}_i$:
\begin{equation}
    A^{3D}_i = A \cdot A^{2D}_i.
\end{equation}
This factorization enforces inter-slab geometric consistency by construction, provided that the shared 3D transformation is topology-preserving (e.g., free of folding). The global transformation captures the overall 3D variability of the reconstruction and ensures that slabs do not intersect or overlap, while the per-slab components account for in-plane variations. Together, these elements yield a coherent 3D reconstruction.

We estimate $A$ and ${A^{2D}_i}$ via alternating solving $A(A^{2D}_i(X_i)) = \hat{Y}_i$ for $A$ and ${A^{2D}_i}$, as described in Alg.~\ref{alg:reconstruction}. 
This procedure yields geometrically coherent reconstructions and naturally generalizes to cases where only a subset of slabs is available, or even a single slab.

\begin{algorithm}[h]
\caption{3D Reconstruction via Alternating Least Squares}
\label{alg:reconstruction}
\KwIn{Predicted MNI maps $\{\hat{Y}_i\}_{i=1}^N$ and 2D pixel grids $\{X_i\}_{i=1}^N$ for $N$ slabs}
\KwOut{Global affine transform $A$, slab-specific 2D transforms $\{A^{2D}_i\}_{i=1}^N$}

Initialize $\{A^{2D}_i\}$ as identity or based on slice index $s_i$\;
\Repeat{convergence or max iterations}{
    \textcolor{Magenta}{\tcp{Update $A$ given $\{A^{2D}_i\}$}}
    \For{$i = 1$ \KwTo $N$}{
        Compute canonical 3D points: $Z_i \gets A^{2D}_i(X_i)$\;
    }
    Stack $Z_i$ and $\hat{Y}_i$ across all $i$\;
    Solve for $A$ such that $A(Z) \approx \hat{Y}$ via least squares\;

    \textcolor{Magenta}{\tcp{Update $\{A^{2D}_i\}$ given $A$}}
    \For{$i = 1$ \KwTo $N$}{
        Solve for $A^{2D}_i$ such that $A(A^{2D}_i(X_i)) \approx \hat{Y}_i$ via least squares\;
    }
}
\Return{$A$, $\{A^{2D}_i\}$}
\end{algorithm}

\begin{figure*}
    \centering
    \includegraphics[width=1\linewidth, trim={0 7cm 0 0}, clip]{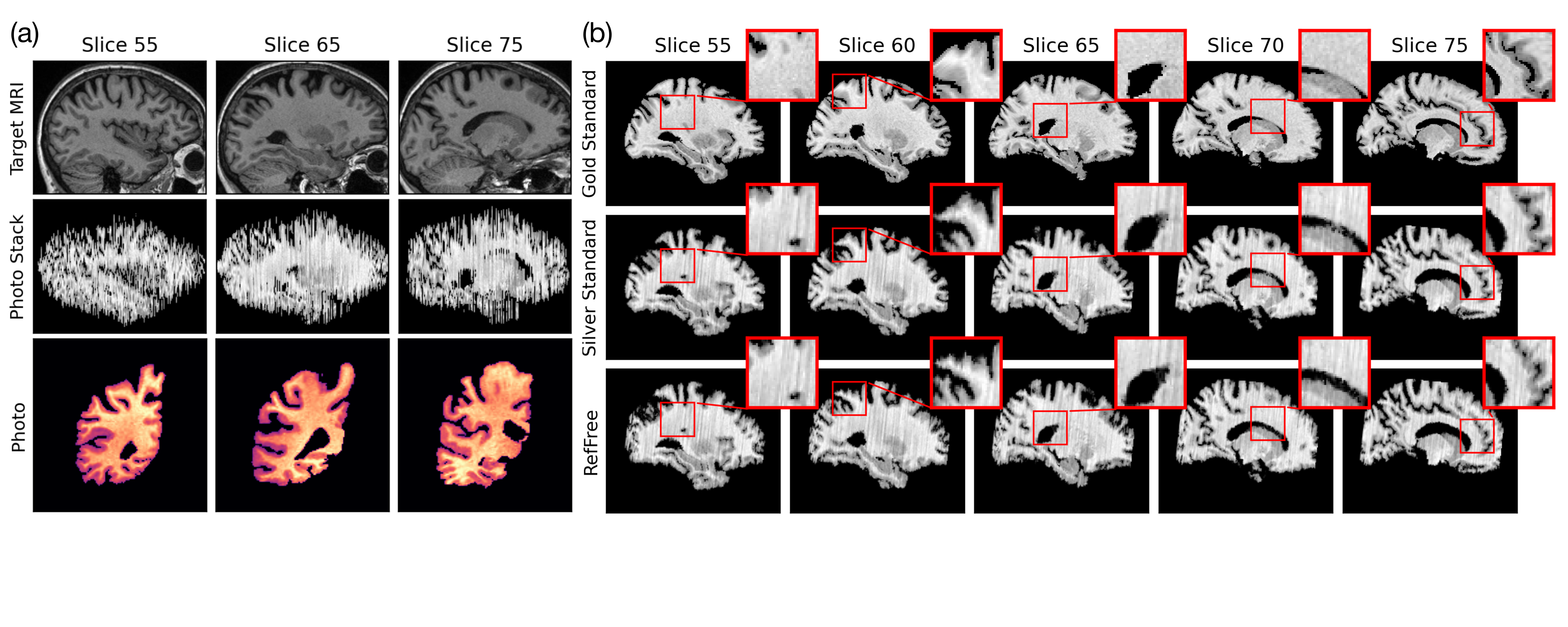}
    \caption{RefFree Reconstruction. (a) The MRI image, simulated photograph stack (sagittal view) and photographs from the same subject; (b) RefFree reconstruction (sagittal view) compared to silver and gold standard given the case shown in (a).}
    \label{fig:simulated_recon_results}
\end{figure*}

\section{Experiments}

\subsection{Datasets}\label{sec:dataset}
We evaluate RefFree qualitatively and quantitatively on real and simulated data:

\paragraph{Composite Brain MRI dataset.} For training, we curated a dataset from 11  brain MRI repositories (ABIDE~\cite{di2014autism}, ADHD200~\cite{brown2012adhd}, ADNI~\cite{jack2008alzheimer,weiner2017alzheimer}, AIBL~\cite{fowler2021fifteen}, FreeSurfer~\cite{fischl2002whole}, COBRE~\cite{mayer2013functional}, Chinese-HCP~\cite{vogt2023chinese}, HCP~\cite{van2012human}, ISBI2015~\cite{carass2017longitudinal}, MCIC~\cite{gollub2013mcic}, OASIS3~\cite{lamontagne2019oasis}), all if which include 1~mm isotropic T1w scans. 
These  were preprocessed with FreeSurfer~\cite{fischl2012freesurfer} as explained in Section~\ref{sec:network} to obtain the training data. After manual quality control of the FreeSurfer outputs, we included  5,279 total subjects. We made a training/validation/test split with 4,217/862/200 scans.

\paragraph{Simulated Dissection Photographs.} We simulate \exvivo photographs for evaluation using the Composite Brain MRI dataset (validation and test sets) following the procedure in~\cite{gazula2024machine} (Fig.~\ref{fig:simulated_recon_results}a). This procedure simulates photographs directly from the MRI. Specifically, it applies random affine to each slice of the MRI volume, followed by applying random illumination to the MRI intensity. We note that this simulation process is different from our proposed synthetic data engine. Using different approaches to generate the \emph{simulated} photographs for evaluation and \emph{synthetic} images for training enables us to mimic the domain gap between training and inference in real scenarios. 

\paragraph{Massachusetts General Hospital Brain Bank}
Dissection photographs of coronal brain slabs were acquired from the Massachusetts Alzheimer's Disease Research Center, affiliated with Massachusetts General Hospital. The dataset includes overhead photographs of coronal slabs from 149 fixed brain specimens. All slabs were manually cut post fixation using an landmark-based protocol that produces thick, uneven slices  ($\sim$10~mm), posing challenges for 3D reconstruction.

\paragraph{University of Washington Fixed Tissue} %
This dataset comprises both \textit{ex vivo} MRI images and dissection photographs of fixed coronal brain slabs from 26 whole-brain specimens sourced at the University of Washington's Department of Laboratory Medicine and Pathology. Following fixation, specimens were embedded in agarose to minimize tissue deformation during slabbing. Prior to dissection, each specimen underwent an \textit{ex vivo} FLAIR MRI scan (at least 1~mm isotropic resolution). Coronal slabs were cut to a consistent thickness of $\sim$4~mm using a modified deli slicer, achieving higher slab uniformity than what is typical in routine neuropathological dissection, and ensuring the best possible 3D reconstructions. 

We train the MNI prediction model on dataset a) and conduct the quantitative evaluation on dataset b), c) and d). Informed consent was obtained for dataset c) and d).

\subsection{Evaluation of Raw MNI Coordinate Prediction}\label{sec:simulated_data}
We evaluate two separate aspects of RefFree: the ability of the U-net to predict MNI coordinates, and the accuracy of the full pipeline, i.e., including the fitting procedure.

\begin{figure*}
    \centering
    \includegraphics[width=\linewidth, trim={0 0.3cm 0 0}, clip]{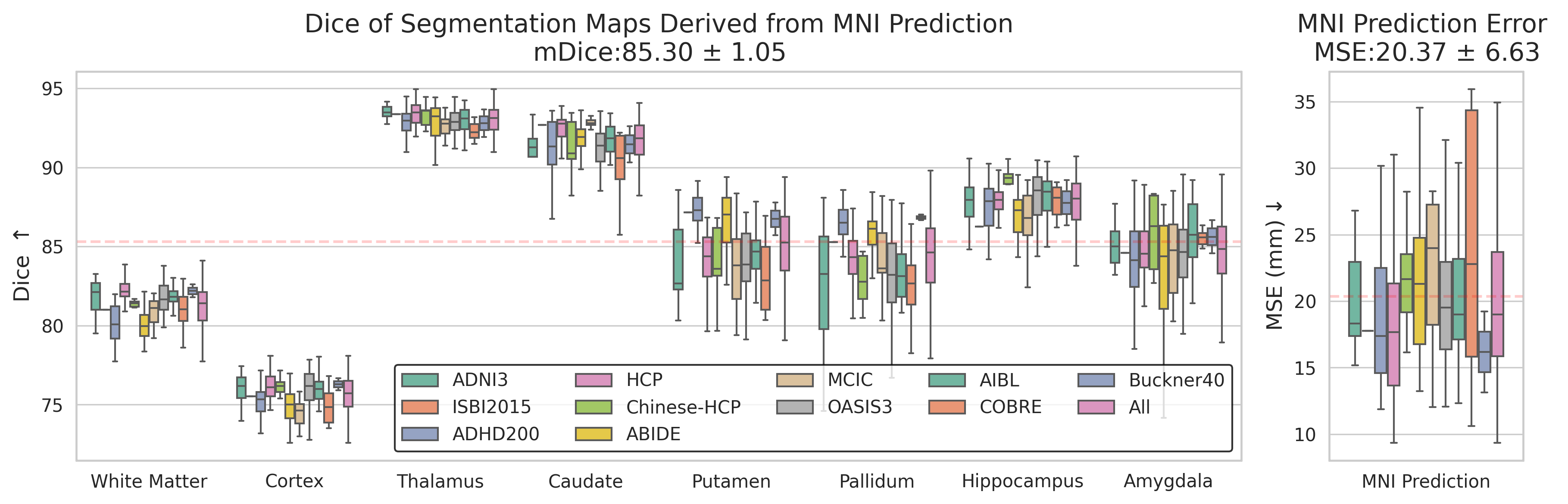}
    \caption{Prediction error that is quantified using MSE and Dice scores. The MSE plots (right) show that the MNI coordinates predicted by RefFree are accurate to within $\sim4.51$ mm on average. The Dice scores (left) between segmentation maps derived from the predicted and ground truth MNI coordinates illustrate how prediction error affects segmentaiton accuracy when using the predicted coordinates for atlas-based segmentation.}
    \label{fig:MNI_prediction_test_set}
\end{figure*}

\subsubsection{MNI Coordinate Prediction}
We compute the mean squared error (MSE) between the predicted and ground truth coordinates, and the Dice score between label maps warped by the predicted and ground truth MNI coordinates as shown in Fig.~\ref{fig:MNI_prediction_test_set}. The overall mean MSE on the test set across datasets is $20.37\pm6.63$, i.e., root MSE under 5~mm, which is quite accurate, particularly when one considers the difficulty of estimating the anterior-posterior ($y$) coordinates out of plane. The average Dice score is $85.30\pm1.05$, computed over 8 representative structures of interest. Fig.~\ref{fig:MNI_prediction_test_set} also shows that the MNI coordinate prediction performs consistently well across highly heterogeneous populations, including healthy subjects (HCP, Chinese-HCP), dementia (OASIS3, ADNI3, AIBL, FreeSurfer), Autism Spectrum Disorder (ABIDE), Attention Deficit Hyperactivity Disorder (ADHD), Schizophrenia (COBRE, MCIC), and Multiple Sclerosis (ISBI2015).

\subsubsection{Ablation Study of the Synthetic Data Engine}
We conduct an ablation study on Steps 1, 3, and 4 and report the MNI coordinate prediction errors in Tab.~\ref{tab:ablation_study}. All components contribute meaningfully to the final performance. The largest improvements arise from the randomized intensity generation and the per-slab deformation simulation. This is expected: without intensity generation, the model would likely overfit to label maps and fail to generalize to photographs; meanwhile, simulating deformation for each slab substantially increases the variability of the synthetic input, which improves robustness to real-world appearance variation.

\begin{table}[h]
    \caption{Ablation study of Steps 1-4: generated intensities, per-slab deformation, 3D affine transform, and croppoing (C for central crop, R for random crop).}
    \label{tab:ablation_study}
    \centering
    \begin{tabularx}{\linewidth}{lcccccc} \toprule
            Model & Intensity & Deform. & Affine & Crop & MSE$\downarrow$ & DICE$\uparrow$\\ \midrule \midrule
            Baseline& & & & & 620.48 & 14.59\\ 
    A& \checkmark& & & & 292.80 & 40.63\\
    B& \checkmark&\checkmark & & & 56.68 & 77.08\\
    C& \checkmark& \checkmark&\checkmark & & 35.21 & 81.00\\
    D& \checkmark& \checkmark&\checkmark & C & 24.74 & 83.95\\
    E& \checkmark&\checkmark & \checkmark& R & 21.69 & 84.88\\\bottomrule
    \end{tabularx}
\end{table}

\subsection{Reconstruction with Full Stack of Photos and Full Pipeline}

\subsubsection{Quantitative Evaluation with Simulated Dissection Photographs}
We adopt two references to evaluate the reconstruction with the MNI prediction algorithm (Sec.~\ref{sec:reconstruction}) on the test set: \textit{(i)}~a silver standard consisting of the volume reconstructed from the ground truth MNI coordinates, and \textit{(ii)}~a gold standard consisting of the MRI used to simulate the photographs. We compute the MSE and Structural Similarity Index Measure (SSIM) between the RefFree reconstruction and both references (Tab.~\ref{tab:simulated_recon_results}, with visual results shown in Fig.~\ref{fig:simulated_recon_results}b). The reconstructed volume closely matches the silver standard (SSIM: 0.930 for the test dataset), while the SSIM with the gold standard (SSIM = 0.889) is lower but similar to that between the silver and gold references (SSIM = 0.892), which serves as a proxy for the upper bound of performance.

\begin{table}[t!]
\caption{RefFree reconstruction accuracy on simulated test dataset. The silver standard is the volume reconstructed by RefFree using ground truth MNI coordinates, while the gold standard is the MRI volume used to simulate the photographs.}
    \label{tab:simulated_recon_results}
    \centering
    \begin{tabularx}{\linewidth}{p{2cm}C{1.2cm}C{1.2cm}C{1.2cm}C{1.2cm}} \toprule 
         &  \multicolumn{2}{c}{Silver Standard} & \multicolumn{2}{c}{Gold Standard}\\ \midrule
         & 
    MSE$\downarrow$&SSIM$\uparrow$& MSE$\downarrow$& SSIM$\uparrow$\\ 
    Initial& 6.45e-2& 0.753& 6.75e-2& 0.758\\ \midrule
    Silver Standard& -& -& 8.31e-3& 0.892\\ 
    RefFree& 4.35e-3& 0.930& 8.51e-3& 0.889\\ \bottomrule
    \end{tabularx}
\end{table}

\subsubsection{Quantitative Evaluation with UW Fixed Tissue Brain Slab Photographs}
We further evaluate RefFree on real \textit{ex vivo} brain slab photographs from the UW Brain Bank. The reconstructed volume is topologically consistent with MRI but not geometrically aligned. To enable comparison, we affinely align both the subject’s MRI and the RefFree reconstruction to the MNI atlas, ensuring evaluation in a common reference frame while preserving inter-structure volumetric relationships. We then apply SynthSeg~\cite{billot2023synthseg} to obtain structure-wise segmentation maps for both the MRI and the reconstructed volume.

As voxelwise overlap metrics (e.g., Dice) are not applicable without spatial alignment between the reconstruction and the MRI, we quantify agreement using two complementary measures. First, we compute the relative volume difference per anatomical structure aggregated across subjects,
$\frac{\left|V^{\text{RefFree}}_{s}-V^{\text{MRI}}_{s}\right|}{V^{\text{MRI}}{s}} \times 100\%$,
where $V^{\text{RefFree}}_{s}$ and $V^{\text{MRI}}_{s}$ denote the structure-wise volumes derived from the RefFree reconstruction and the corresponding MRI, respectively. Second, for each subject, we compute the Pearson correlation across anatomical structures between the RefFree-derived volumes and the MRI-derived volumes, thereby assessing consistency of structure-wise volumetric relationships between the RefFree reconstruction and the true MRI within each subject.

\begin{figure*}
    \centering
    \includegraphics[width=\linewidth, trim={0 22cm 0 0}, clip]{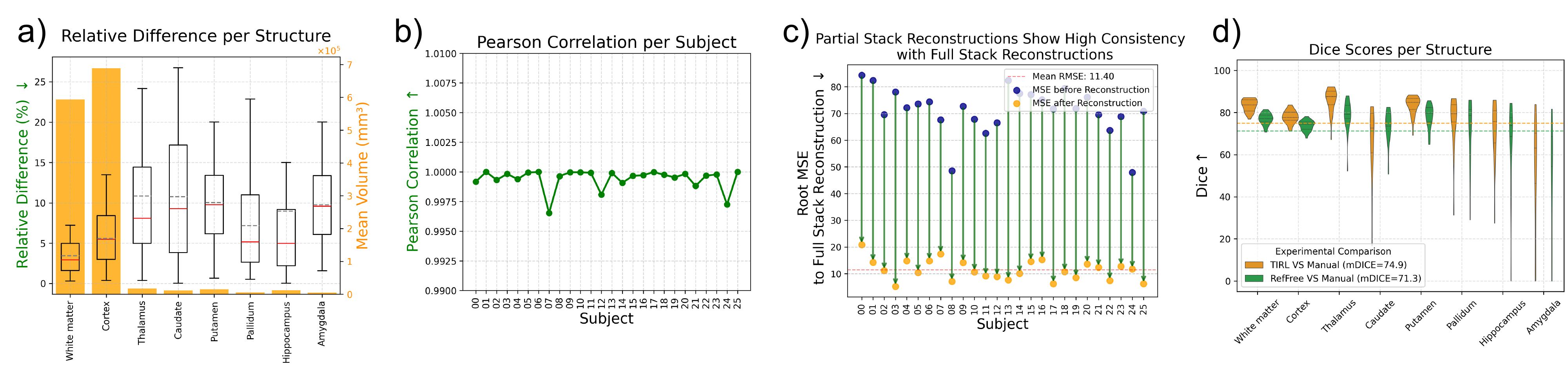}
    \caption{
(a)~Relative volume difference (\%) between RefFree- and MRI-derived volumes per structure (box-and-whisker, left axis) when using full stacks; orange bars show mean absolute structure volume (mm$^3³$, right axis).
(b)~Pearson correlation between MRI- and RefFree-derived volumes (with full stacks).
(c)~Consistency between partial- and full-stack reconstructions. For each subject, 50\% of slab photographs were randomly selected to form a partial stack. MSE was computed between the partial-stack and full-stack reconstructions (orange) and, for comparison, between the raw partial-stack photographs and the full-stack reconstruction (blue).
(d) Dice score between RefFree real-time, atlas-based segmentations and manual label maps on UW fixed-tissue slab photographs.}
    \label{fig:exp_uw}
\end{figure*}

Fig.~\ref{fig:exp_uw}(a) shows box-and-whisker plots of relative volume difference per structure (left axis), with orange bars indicating mean structure volumes (right axis). The median (red) and mean (gray dashed) relative differences fall within $\sim3\%$–$\sim11\%$ across all structures. Because the reconstructed volume and MRI are not spatially aligned, the relative differences are not expected to be zero. The fact that all structures lie in a similar range ($\sim3\%$–$\sim11\%$) indicates that the reconstructed volume preserves inter-structure volumetric relationships. Fig.~\ref{fig:exp_uw}(b) further reports per-subject Pearson correlations, which are uniformly high across the cohort, indicating that RefFree preserves each subject’s inter-structure volumetric profile despite the absence of paired MRI during inference. These findings support that RefFree recovers anatomically consistent volumetric proportions from slab photographs alone under realistic acquisition conditions. %

\subsubsection{Evaluation with MGH Brain Slab Photographs}
We also evaluate RefFree on real \textit{ex vivo} brain slab photographs from the MGH Brain Banks. For these cases, no corresponding MRI scans are available. Therefore, we compare the structure-wise volumes derived from RefFree against those obtained using a reference-based reconstruction baseline~\cite{tregidgo20203d,gazula2024machine}. Specifically, we reconstruct each specimen using the MNI atlas as the reference with the baseline method and then compute, for each subject, the Pearson correlation across anatomical structures between the volumes produced by RefFree and those produced by the baseline method. We report the mean of these per-subject correlations. Quantitative results are summarized in Tab.\ref{tab:real_dataset_comparing_with_opt}, and a qualitative comparison for one specimen is shown in Fig.\ref{fig:recon_real_dataset_exp}-b. The consistently high correlations indicate that RefFree produces volumetrically meaningful reconstructions and serves as a viable alternative when reference scans are not available.

\begin{table*}
\caption{Pearson correlation of the volumes of brain structures, derived from the SynthSeg segmentations of the 3D reconstructions obtained with the optimization-based method in FreeSurfer~\cite{tregidgo20203d} (baseline) and RefFree.}
    \label{tab:real_dataset_comparing_with_opt}
    \centering
    \begin{tabular}{lcccccccc} \toprule
         &  Wh.Ma. & Cortex &Thal. &Caud. &Put. &Pallid. &Hippo. &Amyg.\\ \midrule
         RefFree& 
     0.970& 0.976& 0.968& 0.977& 0.972& 0.967& 0.974&0.960 \\\bottomrule
     \end{tabular}
\end{table*}

\subsection{Reconstruction with a Partial Stack of Photographs}
RefFree supports volume reconstruction from incomplete slab photo stacks. This capability enables reconstructions of cases where only a subset of the slabs is imaged or preserved, which is particularly important for legacy brain bank material. Existing reference-based methods~\cite{tregidgo20203d,gazula2024machine} are not designed to operate under such scenarios.

To evaluate RefFree’s consistency under partial input, we conducted a quantitative experiment using the UW fixed-tissue dataset. For each subject, we randomly selected 50\% of the slab photographs and reconstructed a partial-stack volume. We then compared this reconstruction to the full-stack reconstruction from the same subject by computing the mean squared error (MSE) between the two volumes. As a baseline, we also computed the MSE between the original (unprocessed) partial stack and the full-stack reconstruction.

Fig.~\ref{fig:exp_uw}(c) presents the subject-wise comparison of MSE before and after reconstruction. The results show a substantial reduction in MSE across all subjects, indicating that RefFree significantly improves volumetric consistency when reconstructing from partial inputs. This demonstrates the model’s robustness to missing slices and its ability to recover plausible anatomical structure even under reduced observational coverage. A qualitative example of partial-stack reconstruction is shown in Fig.~\ref{fig:recon_real_dataset_exp}(b).

\subsection{Real-time Atlas-based Segmentation for a Single Slab Photograph}
During brain cutting, real-time segmentation maps projected onto the slab could assist pathologists in locating anatomical structures for tissue sampling (digitally guided dissection), particularly in cortical regions. This is not possible with the existing reference-based methods.

\begin{figure*}
    \centering
    \includegraphics[width=0.95\linewidth, trim={0 7.5cm 2.8cm 0}, clip ]{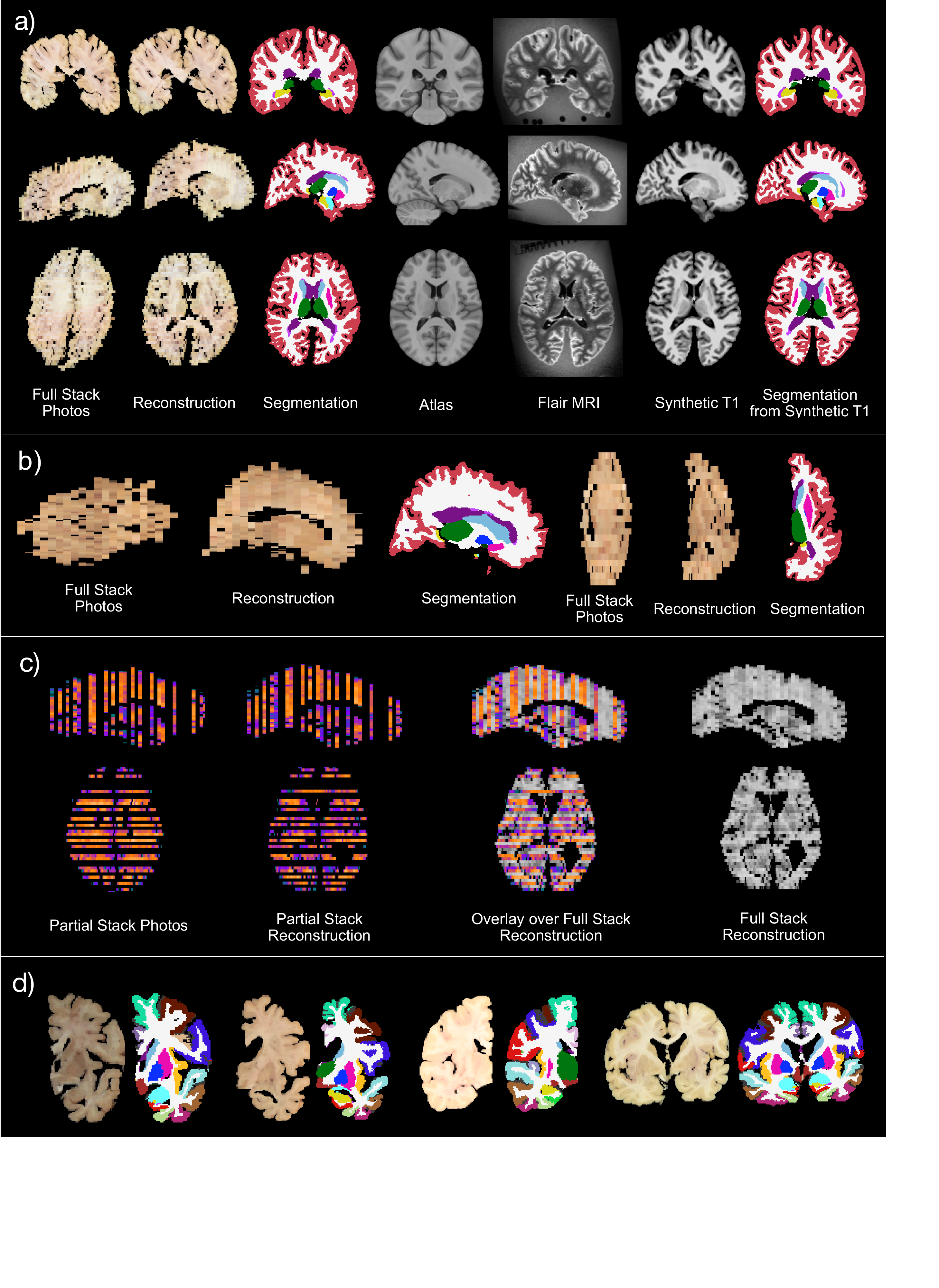}
    \caption{Qualitative results of RefFree on real cases in (a) reconstruction from a full stack of photographs from UW Fixed Tissue dataset, (b) reconstruction from a full stack of photographs from MGH Fixed Tissue dataset, (c) reconstruction from a partial stack, and (d)~real-time atlas-based segmentation of a single photograph on 4 cases from UW Fixed Tissue and MGH Brain Bank datasets.}
    \label{fig:recon_real_dataset_exp}
\end{figure*}

RefFree enables real-time segmentation overlays by projecting atlas labels directly onto individual dissection slab photographs. Given a slab image I with slice index s, the network predicts a per-pixel MNI coordinate map $\hat{Y} : \Omega_s \rightarrow \Omega_{\text{MNI}}$. Atlas labels are then obtained by sampling the atlas segmentation $L_{\text{atlas}}$ via nearest neighbor. This functionality enables real-time digitally guided dissection by providing real-time anatomical context without requiring 3D reconstruction of the full volumes. Fig.~\ref{fig:recon_real_dataset_exp}(c) presents qualitative examples of segmentation overlays for multiple subjects.

To assess accuracy, we evaluate atlas-based segmentations obtained from RefFree against manual annotations drawn on the same slab photographs in the UW fixed-tissue dataset. We further compare the RefFree segmentations with those obtained by registering the photographs to MRI using TIRL~\cite{huszar2023tensor}. The segmentation of the atlas and the MRI are obtained via SynthSeg. Fig.~\ref{fig:exp_uw}(d) reports Dice score per structure aggregated across subjects. RefFree achieves slightly lower Dice score (71.3) than TIRL (74.9) since TIRL performs optimization-based registration directly to the subject’s MRI. Nevertheless, considering that RefFree operates in real time and does not require an MRI, achieving segmentation accuracy close to that of TIRL demonstrates the effectiveness of using RefFree to align a single photograph to the atlas or to estimate segmentation with a single photograph.

\section{Conclusion}
We propose RefFree, a reference-free 3D reconstruction method for dissection photographs trained on \emph{purely synthetic data}. RefFree serves as a viable alternative to the existing reference-based method in FreeSurfer for full-stack reconstruction, while also addressing scenarios where the baseline method falls short (e.g., partial stacks), offering a versatile tool for postmortem brain examination. RefFree also has limitations, such as its reliance on slice indices and its current inability to account for non-rigid deformations during reconstruction. Future work will explore: eliminating the dependency on slice indices; modeling nonlinear deformation, which is crucial when cutting fresh rather than fixed tissue; better simulation of dissection photography using photorealistic rendering; and further validation on downstream applications. In conclusion, this study lays the foundation for robust and generalizable reference-free reconstruction from brain dissection photographs using machine learning.

\bibliographystyle{ieeetr}
\bibliography{reference}

\end{document}